\documentclass[manuscript]{aastex}

\slugcomment{Submitted to the {\it Astrophysical Journal}}

\shorttitle{Thermal Nonequilibrium} \shortauthors{Klimchuk et al.}

\begin{document}


\title{Can Thermal Nonequilibrium Explain Coronal Loops?}


\author{James A. Klimchuk, Judy T. Karpen, and Spiro K. Antiochos}
\affil{NASA Goddard Space Flight Center, Greenbelt, MD 20771}



\begin{abstract}
Any successful model of coronal loops must explain a number of
observed properties.  For warm ($\sim 1$ MK) loops, these include:
1. excess density, 2. flat temperature profile, 3. super-hydrostatic
scale height, 4. unstructured intensity profile, and 5. 1000--5000 s
lifetime.  We examine whether thermal nonequilibrium can reproduce
the observations by performing hydrodynamic simulations based on
steady coronal heating that decreases exponentially with height. We
consider both monolithic and multi-stranded loops.  The simulations
successfully reproduce certain aspects of the observations,
including the excess density, but each of them fails in at least one
critical way. Monolithic models have far too much intensity
structure, while multi-strand models are either too structured or
too long-lived.  Our results appear to rule out the widespread
existence of heating that is both highly concentrated low in the
corona and steady or quasi-steady (slowly varying or impulsive with
a rapid cadence). Active regions would have a very different
appearance if the dominant heating mechanism had these properties.
Thermal nonequilibrium may nonetheless play an important role in
prominences and catastrophic cooling events (e.g., coronal rain)
that occupy a small fraction of the coronal volume. However,
apparent inconsistencies between the models and observations of
cooling events have yet to be understood.
\end{abstract}





\section{Introduction}

It is well known that much of the plasma in the Sun's corona is
confined in distinct loop structures.  The arching shape of these
loops is defined by the magnetic field, but their thermodynamic
properties are determined by the yet-to-be-established mechanism of
coronal heating.  Our understanding of coronal loops and coronal
heating has advanced considerably in recent years, but a number of
important questions remain. We report here on an investigation into
whether ordinary coronal loops can be explained by a phenomenon
known as thermal nonequilibrium. Thermal nonequilibrium occurs
whenever steady or quasi-steady heating is highly concentrated at
low coronal heights in both legs of a loop. It is believed to play
an important role in prominences \citep[e.g.,][]{ak91,kak06}, and it
seems reasonable to consider that its occurrence is more widespread.
Note that quasi-steady heating is here taken to mean heating that
changes slowly compared to a cooling time or that is impulsive but
repeats rapidly compared to a cooling time.

Early observations of coronal loops were made primarily in soft
X-rays and suggested that the loops are in states of static
equilibrium \citep[e.g.,][]{rtv78}.  This implies that the heating
is steady.  Soft X-ray emission is mostly produced by hot ($> 2$ MK)
plasma, but more recent observations made in the extreme ultraviolet
(EUV) have revealed a much different picture at lower temperatures.
Most warm ($\sim 1$ MK) loops are clearly inconsistent with static
equilibrium.  We are referring explicitly to those warm loops that
appear as complete structures within the interiors of active
regions.  We do not consider partial loops, sometimes called ``fan"
loops, that are often seen at the perimeters of active regions.

A number of discrepancies with static equilibrium have been
identified.  Perhaps the most significant concerns the density.
Static equilibrium theory predicts a well defined relationship among
the density, temperature, and length of a loop. Warm loops are
observed to have a much higher density than is expected given the
observed temperature and length \citep{aetal99,asa01,wwm03,k06}. The
density excess is typically a factor of about ten, but factors
ranging from near unity to more than a thousand have been measured.

A second discrepancy between observations and theory concerns the
variation of temperature along the loop.  Observations from
broad-band and narrow-band imagers such as the {\it Transition
Region and Coronal Explorer (TRACE)} can be used to measure
temperature with a method known as the filter ratio technique. The
ratio of intensities obtained in two filters, or band-passes, is
related to a temperature under the assumption that the emitting
plasma is isothermal.  When measured in this way, warm loops tend to
have a temperature profile that is much flatter than expected for
static equilibrium (Lenz et al. 1999; Aschwanden et al. 1999;
Aschwanden, Schrijver, \& Alexander 2001; although see Reale \&
Peres 2000).

A third inconsistency is that the density of warm loops decreases
with height much more slowly than expected for a gravitationally
stratified plasma at the measured temperature. The scale height is
too large by up to a factor of two \citep{asa01}. As a consequence,
loops have a more uniform brightness than static equilibrium would
predict.

There are two additional properties of observed loops that prove
extremely important for constraining the models.  Both are
consistent with static equilibrium.  Most loops do not have
small-scale intensity structure. With occasional exception, there
are no localized bright spots or abrupt transitions in brightness.
This is true for both warm loops \citep{ldk08} and hot loops
\citep{kt96, k00}.

Finally, there is the loop lifetime. Warm loops are typically
visible for 1000--5000 s \citep{wws03, ww05, uww06, uwb09}, though
some can live considerably longer.  Hot loops have a much larger
range of lifetimes, with many persisting for multiple hours
\citep{lkm07}. In all cases the loop lives longer than the cooling
time expected from the measured temperature, density, and loop
length.

Explaining all five of these observed properties is very
challenging. One model that does so successfully postulates that
loops are bundles of unresolved strands that are heated impulsively
by storms of nanoflares; see Klimchuk (2006, 2009) for a discussion
of the basic idea and references to key papers.  In this picture,
each strand is heated once and allowed to cool, but many different
strands are energized over a finite time window, which is the storm
duration. Impulsive heating is very appealing both because it is
able to explain the observations and because all current theories of
heating mechanisms predict that the heating is short lived on
individual magnetic strands \citep{k06}.  This includes wave
heating.  A critical aspect of the nanoflare storm idea is that
strands do not get reheated. The plasma must be allowed to cool from
high temperatures to less than 1 MK in order to explain over-dense
warm loops.  If nanoflares recur in a given strand with a delay that
is significantly shorter than a cooling time, then the conditions
are similar to steady heating.

In the work presented here, we assume that the heating is truly
steady or that the cadence of impulsive heating is sufficiently
rapid that a steady approximation is valid. We further assume that
the heating is highly concentrated near both footpoints of the loop.
Such conditions are known to produce a state of thermal
nonequilibrium
\citep{ak91,skaetal99,ketal01,ketal03,ketal05,mhp03,mph04,kak06,metal08}.
As the name implies, no equilibrium exists. The loop is inherently
dynamic and undergoes periodic convulsions as it searches for a
nonexistent equilibrium. Cold, dense condensations form, slide down
the loop leg, and later reform in a cycle that repeats with periods
of several tens of minutes to a few hours. It has been firmly
established that rapidly repeating, low-altitude nanoflares also
produce a state of thermal nonequilibrium
\citep{tpr05,ka08,setal10,as09}.

Our objective here is to determine whether thermal nonequilibrium
can reproduce the observations described above, in particular the
EUV observations of warm loops: 1. excess density, 2. flat
temperature profile, 3. super-hydrostatic scale height, 4.
unstructured intensity profile, and 5. 1000--5000 s lifetime. Our
approach is to perform numerical loop simulations by solving the 1D
time-dependent hydrodynamic equations. From these, we generate
synthetic data representing observations made in the 171 and 195
channels of {\it TRACE} and the Al.1 and AlMg channels of the Soft
X-ray Telescope (SXT) on {\it Yohkoh}. We then measure temperature
and density using precisely the same filter ratio technique that is
applied to real data.

Our study treats two fundamentally different types of loops.  We
first consider monolithic structures in which the plasma is uniform
over the loop cross section.  We then consider bundles of very thin
unresolved strands, similar to what is envisioned in the nanoflare
storm picture described above.  Just as the nanoflares are assumed
to occur at different times, we assume that the condensation cycles
of thermal nonequilibrium are out of phase in the different strands.
We describe the details of the numerical model in the next section
and present and discuss the simulation results in Sections 3 and 4.

\section{Numerical Model}
Because the solar corona is highly ionized and because the magnetic
field dominates the plasma within active regions (i.e., the plasma
$\beta$ is small), we can treat coronal loop strands as
one-dimensional structures. We therefore perform our numerical
simulations with the Adaptively Refined Godunov Solver (ARGOS) hydro
code \citep{skaetal99}, which solves the 1D hydrodynamic equations
of mass, momentum, and energy conservation:

\begin{equation}
\frac{\partial \rho }{\partial t}  + \frac{\partial}{\partial s}
(\rho\upsilon)=0,
\end{equation}

\begin{equation}
\frac{\partial}{\partial t}(\rho\upsilon) + \frac{\partial}{\partial
s} (\rho{\upsilon}^{2})=\rho g_{\|}  - \frac{\partial P}{\partial
s},
\end{equation}

\begin{equation}
\frac{\partial E}{\partial t} + \frac{\partial}{\partial
s}[(E+P)\upsilon] = \rho\upsilon g_{\|} + \frac{\partial}{\partial
s}(\kappa_{0}T^{5/2}\frac{\partial T}{\partial s}) -n^{2}\Lambda(T)
+ Q,
\end{equation}

\noindent where
\begin{equation}
E=\frac{1}{2}\rho{\upsilon}^{2} +\frac{P}{\gamma-1} .
\end{equation}

\noindent Here, $s$ is the spatial coordinate along the loop; $n$ is
the electron number density; $\rho = 1.67\times{10}^{-24} \times n $
is the mass density assuming a fully ionized hydrogen plasma; $T$ is
the temperature; $P = 2nkT $ is the total pressure; $\upsilon$ is
the bulk velocity; $ \kappa_0 = {10}^{-6}$ is the coefficient of
thermal conduction for Spitzer conductivity; $\gamma = 5/3$ is the
ratio of the specific heats; $g_{\|}$ is the component of gravity
parallel to the loop axis; $Q$ is the volumetric heating rate; and
$\Lambda(T)$ is the optically thin radiative loss function as given
in \citet{kpc08} with the exception that there is a $T^3$ dependence
below 0.1 MK to account approximately for optical depth effects.

ARGOS uses adaptive mesh refinement to dynamically modify the
numerical grid in response to changes in the density gradients. This
is critically important for simulations of this type. The cold
condensations which form and move along the loop are bounded by thin
transition regions similar to the classical transition regions at
the footpoints of loops. Only by subdividing and merging grid cells
is it possible to resolve these structures with a grid of reasonable
size. Our simulations have approximately 3500 total cells while the
condensations are present. The smallest cell length is 6 km.  This
is approximately one-third the temperature scale length in the lower
transition region where $T = 0.1 T_{max}$.

The loop is assumed to be a vertical semi-circle with a
footpoint-to-apex half length of $L=75$ Mm. This half length is
characteristic of warm loops and is the value we have used for some
of our nanoflare studies \citep[e.g.,][]{kpc08}. The cross sectional
area is constant, consistent with observations of both EUV and soft
X-ray loops \citep{k00,lkd06}. Attached to each end of the coronal
semi-circle is a 50 Mm chromosphere/photosphere that is maintained
at a nearly constant temperature of $3\!\times\!10^4$ K by a
radiative loss function that decreases precipitously to zero between
$3\!\times\!10^4$ and $2.95\!\times\!10^4$ K. This loss function
applies to the entire loop, including the cold condensations that
form in the coronal portion. Although the radiative properties of
the chromosphere and condensations are treated in a highly
simplified manner (there is no detailed radiative transfer), the
interaction with the rest of the loop is modeled rigorously. In
particular, the exchange of mass by the important processes of
evaporation and condensation is fully included. Radiative transfer
effects are important for explosive evaporation that is driven by
energetic particle beams penetrating deep into the chromosphere
during flares, but the gentle evaporation in our simulations is due
to a heat flux that is mostly dissipated in the transition region.
Only a small fraction of the heat flux reaches the chromosphere.

We begin each simulation by allowing the loop to relax to a static
equilibrium with a spatially uniform background heating, $Q_b$. The
choice $Q_b = 6\!\times\!10^{-4}$ erg cm$^{-3}$ s$^{-1}$ produces an
apex temperature of 3.0 MK.  Over the next 1000 s, we slowly turn on
a localized heating that decreases exponentially with height above
the chromosphere at both ends of the loop and is spatially uniform
below:
\begin{equation}
Q_l(s \geq s_0) = Q_0 \exp [- (s-s_0)/\lambda]  \label{eq:q_fctn}
\end{equation}
\noindent on the left side, where $s_0 = 50$ Mm is the top of the
chromosphere. The right side is a mirror image with the exception of
amplitude (see below). Both the background and localized heating are
held constant thereafter. The scale length of the exponential
decrease is $\lambda = 5$ Mm, which is 1/15 of the loop half length.
Its maximum amplitude at the left footpoint is $Q_0 = 8.0 \!\times\!
10^{-2}$ erg cm$^{-3}$ s$^{-1}$. We impose an asymmetry by making
the amplitude at the right footpoint only 50, 75, or 90\% as large.
The localized heating provides nearly an order of magnitude more
total energy (spatially integrated over the loop) than does the
uniform background heating, and therefore it dominates.

Our volumetric heating function (Eq. 5) has two broad but crucial
constraints: it must be spatially localized above the chromosphere
with a characteristic scale smaller than 10\% of the loop length,
and quasi-steady in comparison to the ambient radiative cooling
time.  Earlier studies of thermal nonequilibrium, as well as the
physical explanation of thermal nonequilibrium (see below), all
indicate that the basic phenomenon is otherwise independent of the
details of the heating. Therefore, any physical heating mechanism
that satisfies these constraints is capable of producing thermal
nonequilibrium. Identifying which of the many candidates for coronal
heating meet these criteria is an important long-term objective, but
it is beyond the scope of this paper.  Our sole aim is to
investigate whether thermal nonequilibrium can explain ordinary
coronal loops, for which purpose our heating function is
appropriate.

\section{Results}

\subsection{Monolithic Loops}

\begin{figure}
\epsscale{.80}
\plotone{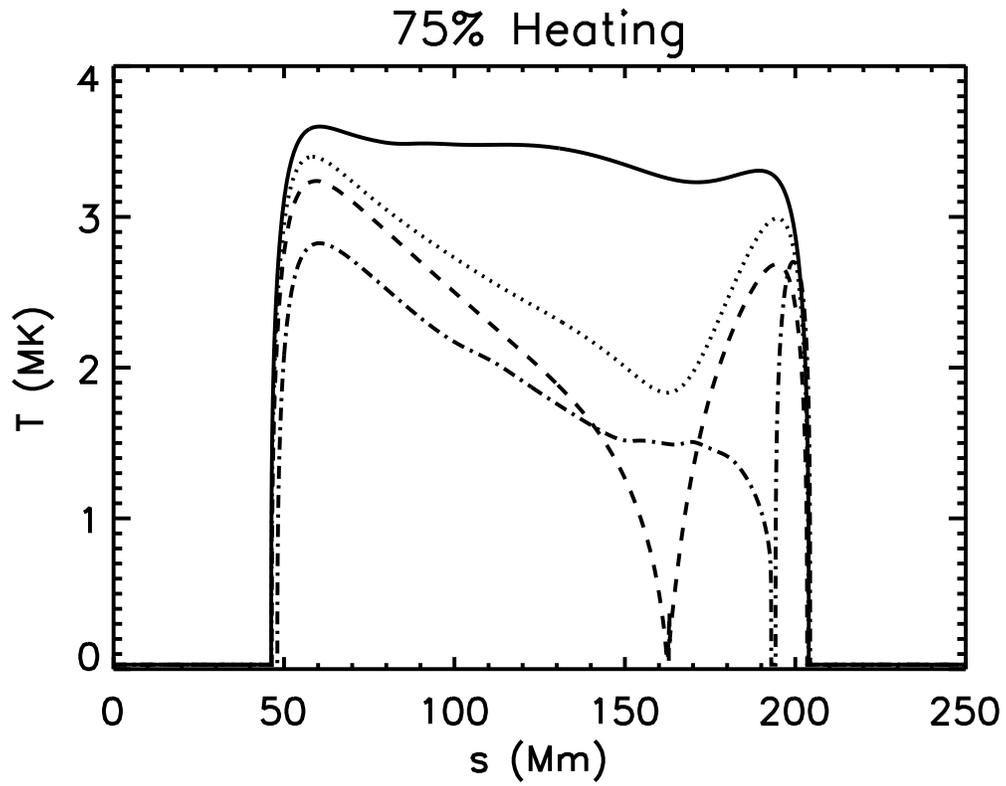}
\caption{Temperature versus position
at four times during the sixth condensation cycle of the loop with
strong heating and 75\% asymmetry: $t$ = 2950 (solid), 4500
(dotted), 4850 (dashed), and 5750 (dot-dashed) seconds after the
previous condensation hits the chromosphere.
\label{fig:hot_75_temp}}
\end{figure}

The first loop we consider is monolithic and has a 75\% heating
asymmetry. It exhibits quasi-periodic behavior with condensations
forming roughly every 6000 s.  Figure \ref{fig:hot_75_temp} shows
the temperature profile at four different times during the sixth
condensation cycle, long after any memory of the initial static
conditions has disappeared. The evolution is typical of thermal
nonequilibrium and has been well documented in our other work. After
the condensation from the fifth cycle falls to the chromosphere ($t
= 0$ s), the loop rapidly heats and attempts to establish an
equilibrium. A peak temperature of 4.4 MK is reached at $t = 650$ s.
This is followed by a relatively long period of slow cooling.  The
solid curve in Figure \ref{fig:hot_75_temp} shows the temperature
profile at $t = 2950$ s, well into the cooling phase.

The reason for the slow cooling and eventual formation of a
condensation can be understood as follows.  We begin by noticing
that the maximum temperature $T_{max}$ occurs close to the left
footpoint, at a height comparable to the heating scale length
$\lambda$. Let us hypothetically divide the loop into two unequal
parts: a short section to the left of $T_{max}$ and a much longer
section to the right. Imagine that the short section is one-half of
a small symmetric loop.  If this hypothetical loop were in static
equilibrium, it would satisfy the scaling law

\begin{equation}
n = 1.32\!\times\!10^6\, \frac{T_{max}^2}{\ell} ,  \label{eq:n_law}
\end{equation}

\noindent where $\ell$ is the half length, approximately equal to
$\lambda$. Equation (\ref{eq:n_law}) follows from the well-known
scaling law $T_{max} = 1.4\!\times\!10^3\, (P\ell)^{1/3}$
\citep{rtv78} upon substituting for $P$ using the ideal gas law. The
downward heat flux from $T_{max}$ is very large due to the steep
temperature gradient. Correspondingly large density is required in
order for radiation from the transition region to balance the heat
flux.  If the actual density is smaller than the equilibrium value
given in equation (\ref{eq:n_law}), the radiation will be too weak,
and chromospheric evaporation will occur, as it does in our
simulation.

Now consider the other section of the original loop, to the right of
$T_{max}$. Imagine that it is half of a different hypothetical loop.
It has the same maximum temperature as the short loop, but because
it is much longer, its equilibrium density is much smaller according
to equation (\ref{eq:n_law}). Of course the long and short ``loops"
are really attached.  Evaporation in the short section drives up the
density in the long section to values that exceed the local
equilibrium conditions. Radiation is enhanced at the elevated
densities, so the plasma cools.

The above argument based on the static equilibrium theory shows why
static conditions are not possible with highly localized footpoint
heating, but the actual energy balance in the evolving loop is more
involved due to the presence of flows. The evaporating material
carries an enthalpy flux that plays a very important role. It
provides nearly enough energy to power the enhanced coronal
radiation. This is the reason why the evolution is so slow during
most of the cooling phase. In fact, if only the left leg were
subjected to localized heating, the loop would establish a dynamic
equilibrium with a steady end-to-end flow and no cooling
\citep{pkm04}. Our loop has localized heating on both sides, which
drives evaporative upflows from both ends. Because material
continually accumulates in the corona, the plasma must cool, and no
steady state is possible.

We see from Figure \ref{fig:hot_75_temp} that the cooling is not
symmetric.  Because evaporation is stronger on the left side than on
the right, the flows converge to the right of the loop midpoint.
Cooling is fastest at this location, and a dip develops in the
temperature profile. The dip grows at an accelerating pace until a
cold condensation is ultimately formed at $t = 4850$ s (dashed
curve). The final collapse resembles a thermal instability; only 350
s are required for the temperature to drop from 2.0 to 0.03 MK. Once
formed, the condensation is pushed to the right by a small pressure
imbalance. It hits the chromosphere approximately 1300 s after first
appearing, and a new condensation cycle begins.

\subsubsection{Excess Density Factor}

The model loop is characterized by over-dense conditions during most
of its evolution. We wish to make a quantitative comparison with
observations, and because many studies of observed loops involve
spatial and temporal averages, we define an excess density factor
$\Psi$ as follows:

\begin{equation}
\Psi = \frac{\bar{n}}{\bar{n}_{eq}} , \label{eq:psi}
\end{equation}

\noindent where
\begin{equation}
\bar{n}_{eq} = 1.32\!\times\!10^6\, \frac{\bar{T}^2}{L}
\label{eq:n_eq}
\end{equation}

\noindent and $\bar{n}$ and $\bar{T}$ are the density and
temperature averaged over the upper 50\% of the loop and over one or
more condensation cycles. Equation (\ref{eq:n_eq}) comes from the
\citet{rtv78} scaling law, analogous to equation (\ref{eq:n_law}).
Averaging over the 11 cycles of our simulation gives $\Psi = 4.09$.
Note that these are simple averages using densities and temperatures
taken directly from the simulation output.  The excess density
factor obtained in this way is different from what we would get from
observed intensities, which provide nonlinear averages of density
and temperature. Later we will perform a more rigorous analysis that
takes this into account.

Because many observed loops are shorter or longer than our model
loop, it is important to examine how $\Psi$ depends on loop length.
We therefore consider two additional models that are half and twice
as long as the original: $L = 37.5$ and 150 Mm. The heating scale
length $\lambda$ and 75\% asymmetry are the same as before, but we
modify the heating amplitudes $Q_b$ and $Q_0$ so that the peak
temperatures of the initial equilibrium and of the condensation
cycles are similar in all three cases. The resulting excess density
factors are 2.90 and 6.62 for the short and long loops,
respectively.

These three cases suggest the relationship $\Psi \propto L^{1/2}$.
We can understand the square-root dependence by considering the
period of slow cooling that dominates the evolution. As discussed
above, a strong downward heat flux drives an upward enthalpy flux in
the lower legs:

\begin{equation}
\kappa_0 \frac{T_{max}^{7/2}}{\lambda} \approx \frac{5}{2} P v .
\label{eq:law1}
\end{equation}

\noindent The enthalpy is nearly enough to power the radiative
losses from the rest of the loop:

\begin{equation}
\frac{5}{2} P v \approx n^2 \Lambda(T_{max}) L . \label{eq:law2}
\end{equation}

\noindent Combining, we get

\begin{equation}
n \approx \left[\frac{\kappa_0
T_{max}^{7/2}}{\Lambda(T_{max})}\frac{1}{\lambda L}\right]^{1/2}
\label{eq:law3}
\end{equation}

\noindent for the actual loop density.

In static equilibrium, the energy loss rates from radiation and
thermal conduction are approximately equal in the corona
\citep{vau79}:

\begin{equation}
n_{eq}^2 \Lambda(T_{max}) \approx \kappa_0 \frac{T_{max}^{7/2}}{L^2}
. \label{eq:law4}
\end{equation}

\noindent This gives

\begin{equation}
n_{eq} \approx \left[\kappa_0
\frac{T_{max}^{7/2}}{\Lambda(T_{max})}\frac{1}{L^2}\right]^{1/2} ,
\label{eq:law5}
\end{equation}

\noindent for the equilibrium density corresponding to $T_{max}$ and
$L$.\footnote{Comparing equations \ref{eq:law5} and \ref{eq:n_eq},
we see that the Rosner, Tucker, \& Vaiana scaling law uses
$\Lambda(T) \propto T^{-1/2}$.} The excess density factor is
therefore
\begin{equation}
\Psi = \frac{n}{n_{eq}} \approx \left(\frac{L}{\lambda}\right)^{1/2}
. \label{eq:law6}
\end{equation}

\noindent  Note that it depends not on the loop length alone, but on
the ratio of the loop length to heating scale length. In principle,
we could reproduce model loops with any $L$ and $\Psi$ simply by
adjusting the value of $\lambda$.  It seems, therefore, that the
observed excess densities of warm loops can be readily explained
with thermal nonequilibrium.

\subsubsection{Intensity}

A successful loop model must also explain the intensity properties
of observed loops, both temporal and spatial. We therefore generate
light curves and intensity profiles for simulated {\it TRACE}
observations of the models made in the 171 channel. We assume that
the loops are viewed from the side, so the intensity at any point
along the loop axis is given by $I = n^2 G(T)$. Here, $G(T)$ is the
instrument response function, which for the 171 channel is
reasonably sharply peaked near 1 MK. We have ignored the loop
diameter and a possible filling factor because we are concerned only
with relative intensities, and both the diameter and filling factor
are assumed to be constant along the loop and unchanging in time.

\begin{figure}
\epsscale{.80} \plotone{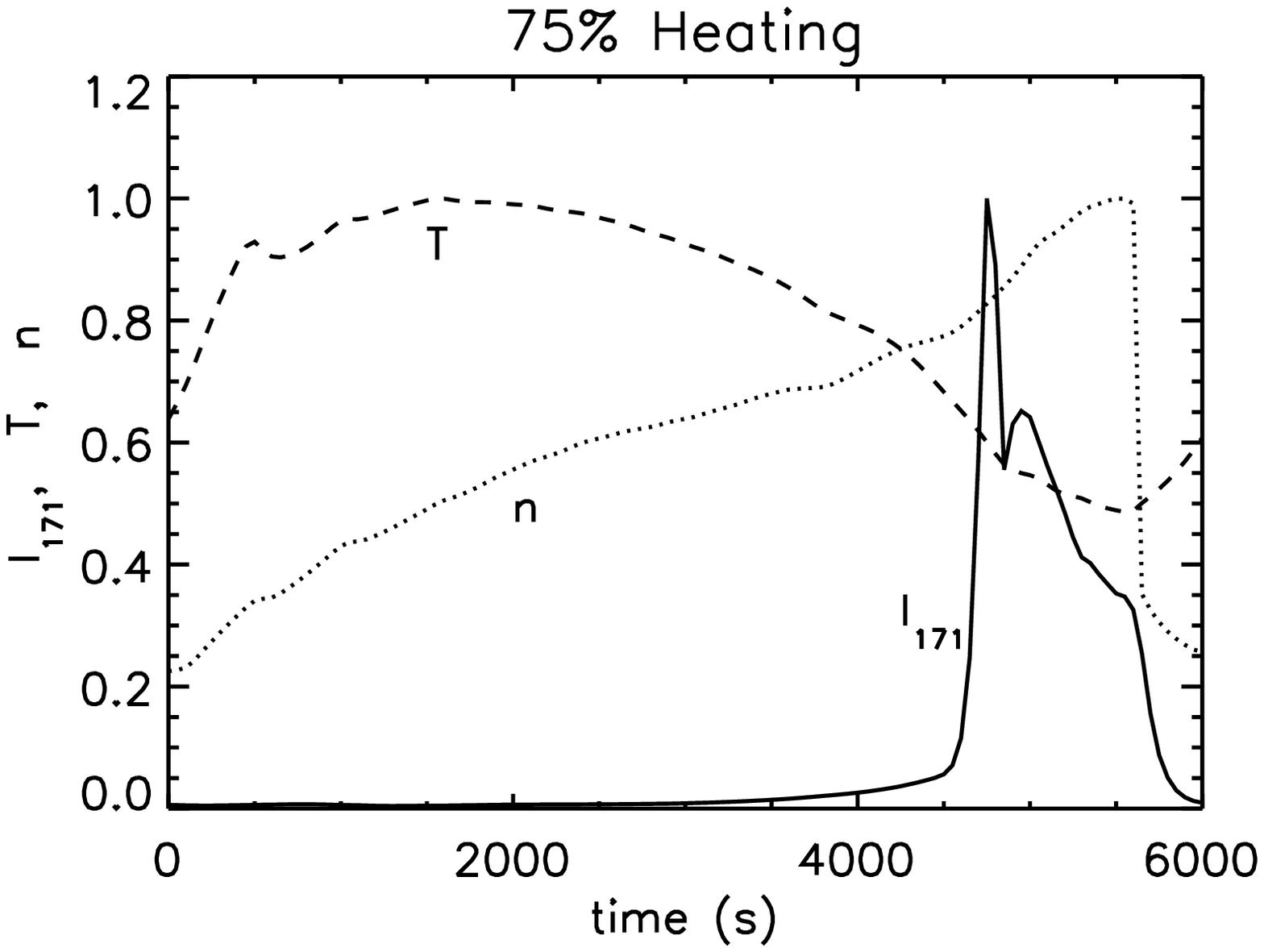} \caption{{\it TRACE} 171 intensity
(solid), temperature (dashed), and density (dotted) versus time for
the sixth condensation cycle of the loop with strong heating and
75\% asymmetry. Values are averaged over the upper 80\% of the loop
and are normalized to their respective maxima.
\label{fig:hot_75_evol}}
\end{figure}

The solid curve in Figure \ref{fig:hot_75_evol} is the light curve
for the sixth condensation cycle of the original $L = 75$ Mm loop.
This is the same condensation cycle shown in Figure
\ref{fig:hot_75_temp}. We have averaged the intensity over the upper
80\% of the loop to exclude the ``moss" emission from the transition
regions at the footpoints.  The dashed and dotted curves show the
corresponding evolution of the spatially averaged temperature and
density.  It is readily apparent how evaporation slowly fills the
loop with plasma.

Before the condensation forms, the coronal plasma is too hot to be
easily detected in the 171 channel, and the light curve is extremely
faint. It brightens dramatically when the condensing plasma cools
rapidly through 1 MK (sharp spike at $t = 4700$ s).  This contrasts
with the much more gradual brightening exhibited by most observed
loops \citep{wws03}. The light curve remains bright after the
condensation has fully formed because transition regions are present
on either side of the cold mass. After about 1000 s the light curve
suddenly dims as the condensation moves out of the 80\% averaging
window. The spatially-averaged density drops at the same time since
the condensation contains most of the loop's mass. Bright emission
is actually present in the loop for another 300 s as the
condensation traverses the remaining 20\% of the leg before hitting
the chromosphere.  The total lifetime in 171 emission is therefore
approximately 1300 s. This is at the extreme low end of the range of
observed lifetimes.

\begin{figure}
\epsscale{.80} \plotone{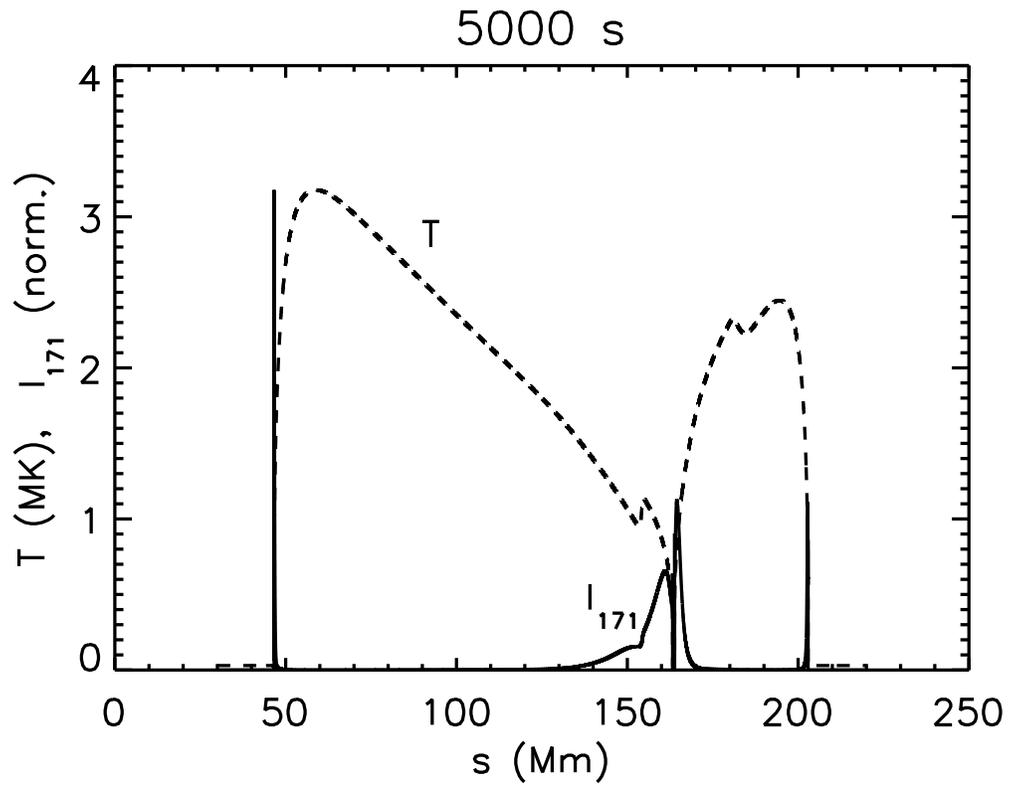} \caption{{\it TRACE} 171 intensity
(solid) and temperature (dashed) versus position at $t = 5000$ s in
the sixth condensation cycle of the loop with strong heating and
75\% asymmetry. \label{fig:hot_75_profile}}
\end{figure}

The abrupt appearance and disappearance of the 171 emission
disagrees with observations, which show a more gradually evolving
light curve.  The spatial distribution of the emission presents an
even bigger problem. Figure \ref{fig:hot_75_profile} shows profiles
of intensity (solid) and temperature (dashed) at $t = 5000$ s, after
the condensation has formed. The emission is highly concentrated in
transition region layers at the loop footpoints (``moss") and on
either side of the condensation. This contrasts sharply with
observed loops, which tend to have a far more uniform appearance.
Falling bright knots are sometimes observed, but these are only
detected at much cooler temperatures ($\leq 0.1$ MK)
\citep{s01,detal04,obd07}. We return to the subject of these knots
later in the Discussion section.

To determine whether the extreme nonuniformity in the intensity
distribution is affected by the degree of heating asymmetry, we
perform two additional simulations using the same heating amplitude
and scale length as before, but with asymmetries of 50\% and 90\%.
The results are qualitatively similar to the 75\% case. The
intensity profiles are highly structured and in gross disagreement
with observations.

The primary reason for the nonuniform intensity is that most of the
loop is too hot to be easily detected in the 171 channel (i.e.,
significantly hotter than 1 MK). To obtain temperatures more
suitable to 171, we perform three new simulations with the heating
amplitude reduced by an order of magnitude.  All other parameters
are as before.  The model with 75\% asymmetry reaches a maximum
temperature of 1.8 MK and has an excess density factor $\Psi =
4.69$. The results for the 50\% and 90\% cases are similar.

\begin{figure}
\epsscale{.80} \plotone{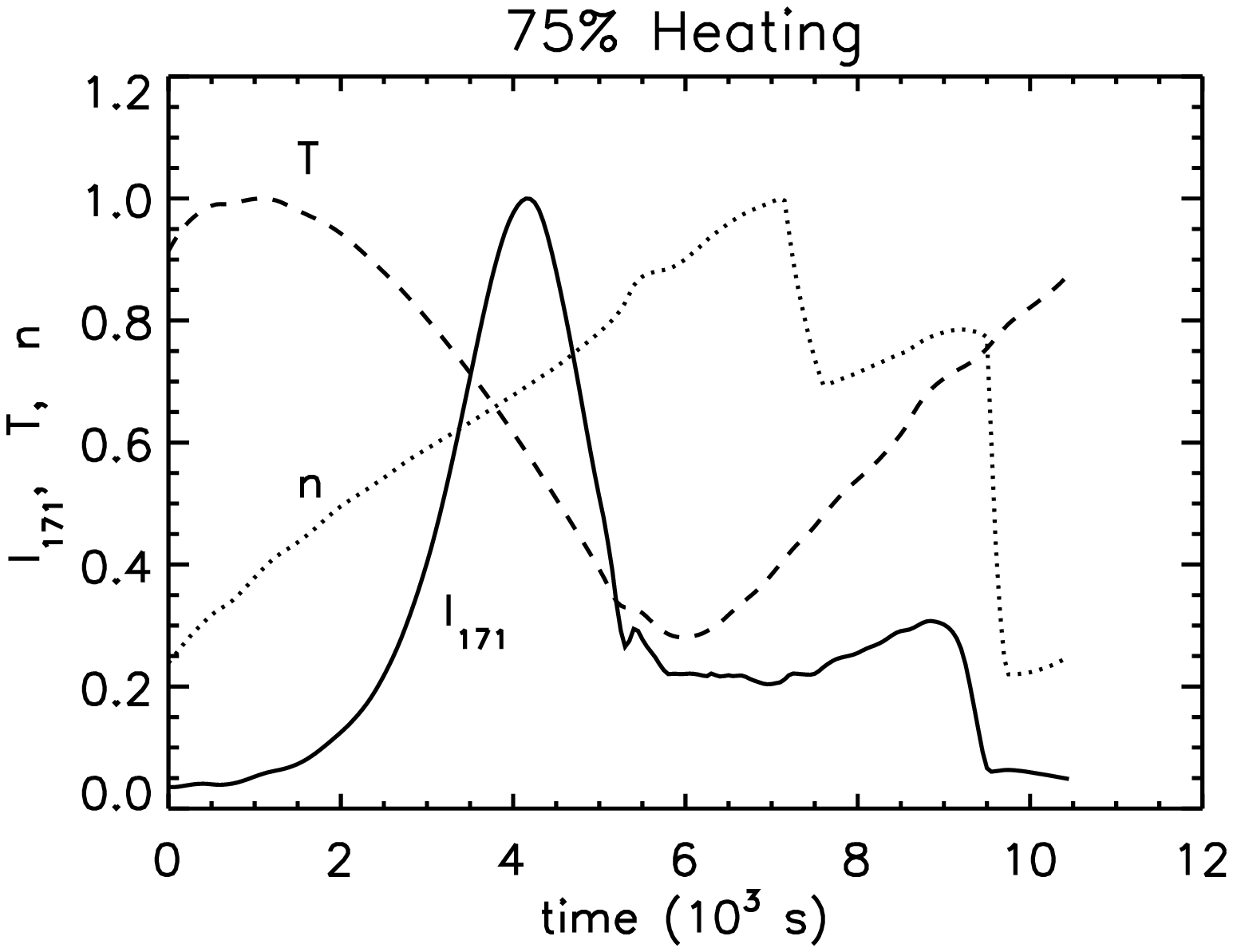} \caption{{\it TRACE} 171 intensity
(solid), temperature (dashed), and density (dotted) versus time for
the second condensation cycle of the loop with weak heating and 75\%
asymmetry. Values are averaged over the upper 80\% of the loop and
are normalized to their respective maxima. \label{fig:warm_75_evol}}
\end{figure}

Figure \ref{fig:warm_75_evol} shows the 171 light curve and the
temperature and density evolution for the second condensation cycle
of the 75\% case. The cycle lasts approximately 11,000 s, nearly
twice as long as the strong heating counterpart. The light curve has
three rather distinct phases---faint, bright, and
intermediate---which is not consistent with the slowly varying
intensities of most observed warm loops. The bright and intermediate
phases together last about 7000 s, which is longer than most
observed loop lifetimes.

\begin{figure}
\epsscale{.80} \plotone{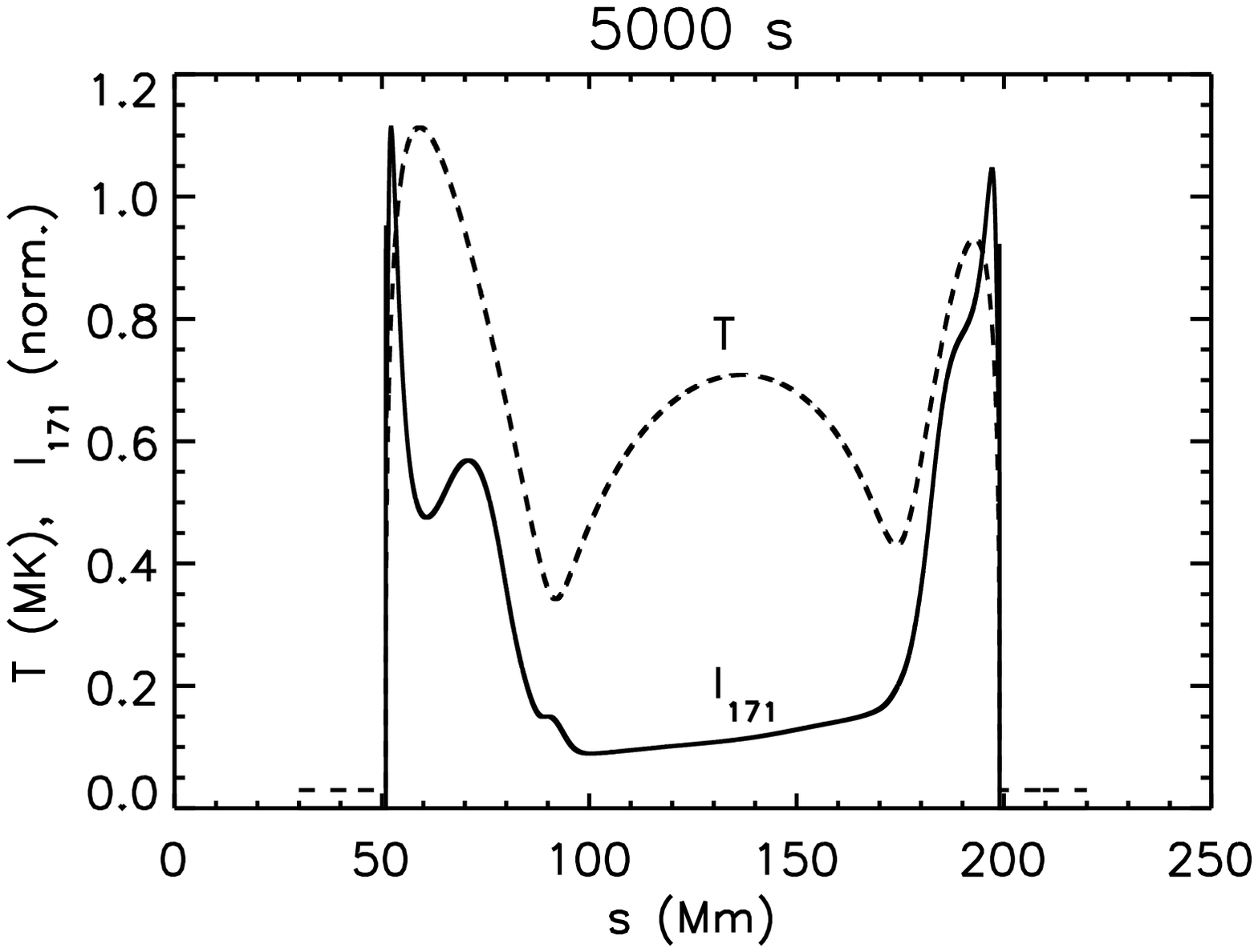} \caption{{\it TRACE} 171 intensity
(solid) and temperature (dashed) versus position at $t = 5000$ s in
the second condensation cycle of the loop with weak heating and 75\%
asymmetry. \label{fig:warm_75_profile1}}
\end{figure}

\begin{figure}
\epsscale{.80} \plotone{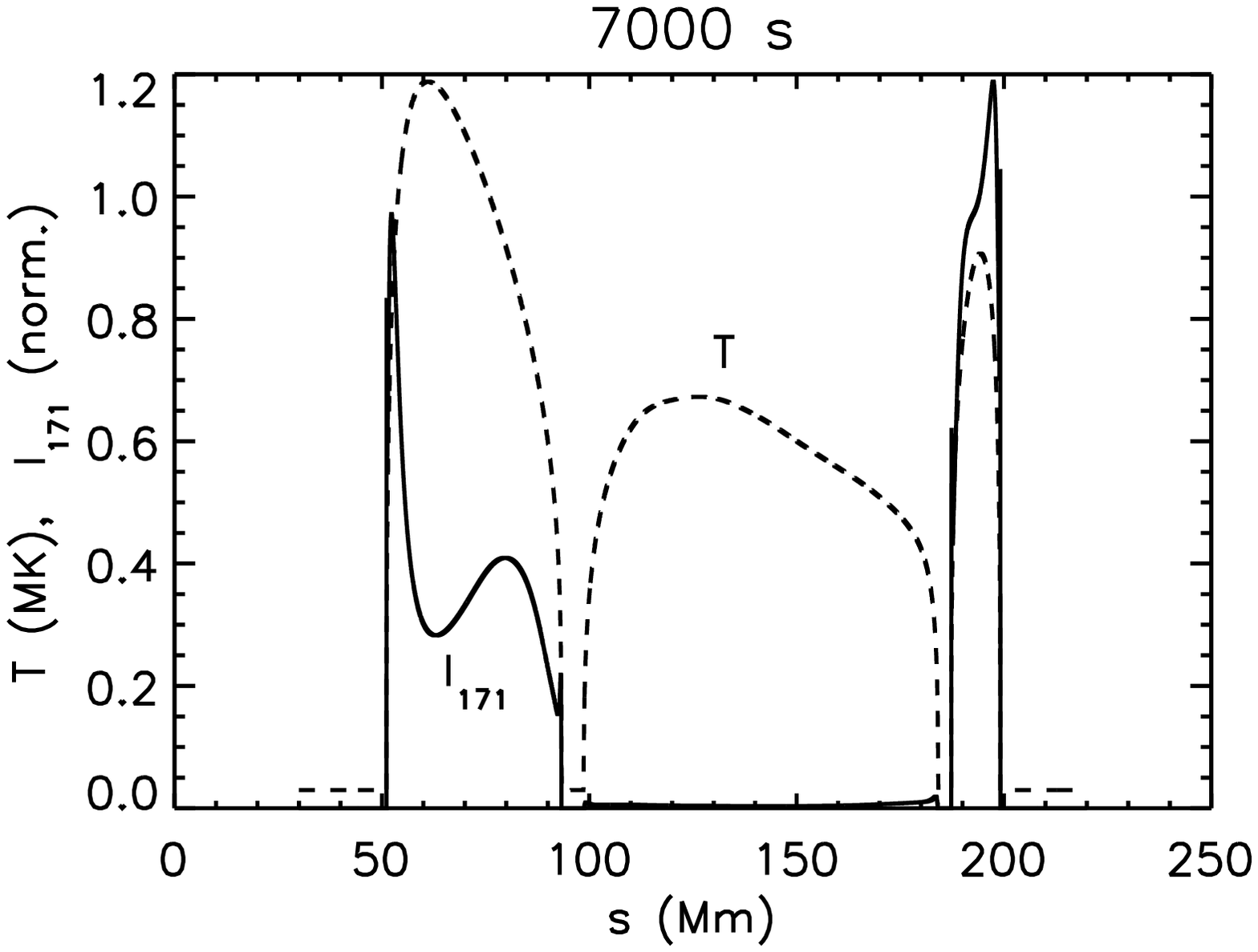} \caption{{\it TRACE} 171 intensity
(solid) and temperature (dashed) versus position at $t = 7000$ s in
the second condensation cycle of the loop with weak heating and 75\%
asymmetry. \label{fig:warm_75_profile2}}
\end{figure}

An interesting aspect of this simulation is that two condensations
are present at the same time, as was seen in earlier studies
\citep{mph04,ketal05}. Figures \ref{fig:warm_75_profile1} and
\ref{fig:warm_75_profile2} show intensity and temperature profiles
at $t=5000$ and 7000 s, before and after the condensations form.
More of the loop is visible than in the strong heating models, but
the intensity still is far more structured than is observed. In
particular, the region between the condensations is extremely faint.
We can understand this behavior as follows. When the condensations
form, the central region between them is cut off from the
evaporative upflows and associated enthalpy flux that powers the
radiative losses. The plasma cools and drains onto the
condensations.  The condensations behave like chromospheres, and a
quasi-static loop equilibrium is established between them. Because
the heating rate is so small, the equilibrium state has a low
temperature and very low density, so the 171 emission is minimal.
The precise value of the temperature and density depend on the
magnitude of the uniform background heating, which dominates in this
part of the loop.  Note that the two condensations remain separate
at all times and do not merge, as is sometimes seen in other
simulations \citep[e.g.,][]{ketal05}.

Thermal nonequilibrium clearly cannot explain observed loops if the
loops are monolithic structures, at least not with steady,
exponential heating of the type we have considered.

\subsection{Multi-Strand Loops}

\subsubsection{Excess Density Factor}

Because our monolithic models fail, we now consider loops that are
bundles of many unresolved strands. To start, we assume that all of
the strands in a given loop are identical except for the phasing of
the condensation cycles, which we take to be random.  We can then
approximate an instantaneous snapshot of the composite loop by
simply time averaging one simulation over one or more cycles.

A wide variety of temperatures coexist within the cross section of
such a multi-stranded loop. The single temperature that is measured
by an instrument like {\it TRACE} or SXT/{\it Yohkoh} is a weighted
average, where the weighting depends on both the temperature
response function, $G(T)$, and the differential emission measure
distribution, $DEM(T)$. To simulate realistic measurements from our
models, we first compute intensity profiles for the individual
strands (i.e., for each time in the simulation), and then we average
them together to obtain a single intensity profile for the loop
bundle.  We do this separately for the 171 and 195 channels of {\it
TRACE} and the Al.1 and AlMg channels of SXT. We next infer
temperature and emission measure, $EM$, at each position along the
loop using both the 171/195 and Al.1/AlMg ratios. From the emission
measure, we compute density according to
\begin{equation}
n = \left(\frac{EM}{d f}\right)^{1/2} ,  \label{eq:density}
\end{equation}
where $d$ is the loop diameter and $f$ is the filling factor.  The
diameter plays no role, since $EM$ is derived from the loop
intensity, which scales with the assumed diameter.  We take a
filling factor of unity, precisely as done for real data, which
means that the density given by equation (\ref{eq:density}) is a
lower limit. Finally, we average $T$ and $n$ along the upper 50\% of
the loop \footnote{The reader may wonder why we use 50\% averages
here and 80\% averages for the light curves.  50\% was used in the
observational studies of loop density to which we will compare our
models. For the light curves, we are only concerned with excluding
the bright moss emission at the footpoints.} and use equations
(\ref{eq:psi}) and (\ref{eq:n_eq}) to obtain the excess density
factors that would be measured by {\it TRACE} and SXT, designated
$\Psi_{TRACE}$ and $\Psi_{SXT}$. We follow this procedure separately
for all 6 of the $L = 75$ Mm models (2 heating amplitudes and 3
heating asymmetries).

It seems unlikely that all of the strands in a given loop bundle
would be identical except for their phases. Therefore, we also build
a composite loop with strong heating and a composite loop with weak
heating by averaging together the models with 50, 75, and 90\%
asymmetry. The averages include both the original models, which have
greater heating in the left leg, and their mirror images, which have
greater heating in the right leg. The two composite loops so
obtained have a mixture of strands of different types, which is
perhaps more realistic. We simulate temperature and density
measurements of these loops using the same procedure described
above, first averaging the intensities and then applying the filter
ratio technique.

Results for the ``homogeneous" multi-strand models and the composite
multi-strand models are presented in Table 1. The first column gives
the loop half length, which is the same for all except the last two
cases. The second column gives the amplitude of the localized
heating together with an indication of whether it is strong
(produces a peak temperature near 4.4 MK) or weak (produces a peak
temperature near 1.8 MK).  The third column gives the heating
asymmetry. The fourth column gives the number of condensation cycles
used in the temporal averages. The fifth column gives the average
period of the cycles, which are only quasi-regular in most cases.
The sixth and seventh columns give the temperatures that would be
measured with {\it TRACE} and SXT filter ratios. The last three
columns give the excess density factors obtained directly from the
temperatures and densities of the models, equation (\ref{eq:psi}),
and from the simulated {\it TRACE} and SXT measurements. The values
differ because {\it TRACE} preferentially detects the warm plasma
and SXT preferentially detects the hot plasma.

\begin{figure}
\epsscale{.80} \plotone{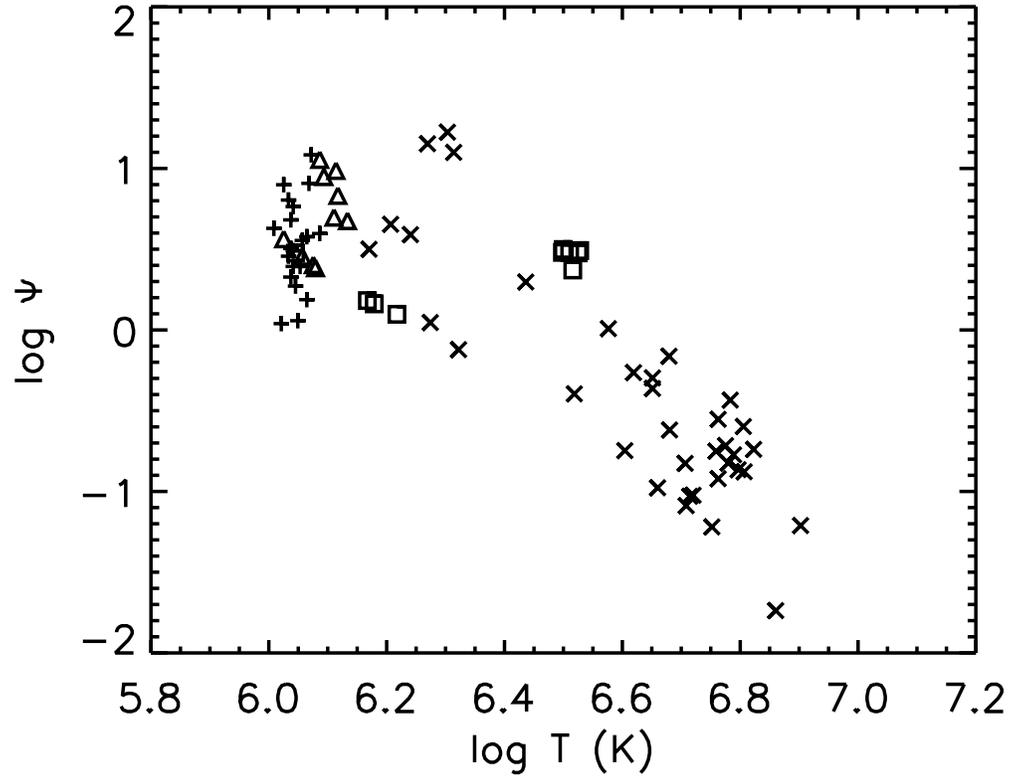} \caption{Excess density factor
versus temperature for real loops observed by {\it TRACE} (pluses)
and SXT (crosses) and for model loops with simulated observations by
{\it TRACE} (diamonds) and SXT (squares). \label{fig:nfact_obs}}
\end{figure}

Figure \ref{fig:nfact_obs} shows the excess density factors of real
loops plotted against temperature.  The factors were determined
precisely as described above, i.e., using equation
(\ref{eq:density}). The loops near 1 MK (pluses) were observed by
{\it TRACE} and analyzed originally by \citet{ana00}. The hotter
loops (crosses) were observed by SXT and analyzed originally by
\citet{pk95} \citep[also][]{kp95}. These are the same loops
presented in Figure 4 of \citet{wwm03} and Figure 6 of \citet{k06}.
Also shown are the excess density factors of the model loops as
determined from simulated {\it TRACE} observations (diamonds) and
simulated SXT observations (squares).

There is good agreement between the models and observations for the
excess density factors obtained from {\it TRACE}. Values range
between about 3 and 11 for the models and between about 1 and 12 for
the observations (note that logarithms are plotted in the figure).
The temperatures measured by {\it TRACE} are also in good agreement,
but this is expected because the 171 and 195 filters have a narrow
temperature response and are only sensitive to plasma close to 1 MK.

The agreement between the models and observations is much worse for
the SXT measurements. Excess density factors from the models are
tightly clustered between 1 (no excess) and 3, whereas those from
the observations range all the way from 0.02 (highly under dense) to
16. The agreement is better if we restrict ourselves to the
temperature range of the models ($1.1 < T_{SXT} < 3.4$ MK), in which
case the observed excess density factors are all $> 0.4$ (slightly
under dense).  However, the observed factors have a strong tendency
to decrease with temperature, while the model factors have a weak
tendency to increase.  We conclude that the models are consistent
with at most a subset of observed SXT loops. According to equation
(\ref{eq:law6}), thermal nonequilibrium can never produce the
under-dense conditions observed at high temperatures because
$\lambda > L$ gives rise to static equilibrium (in fact, static
equilibrium occurs whenever $\lambda > L/5$ approximately). It is
worth pointing out that the nanoflare storm model is capable of
explaining both over-dense warm loops and under-dense hot loops
\citep{k06}.

The quantity $\Psi$ defined in equations (\ref{eq:psi}) and
(\ref{eq:n_eq}) very likely underestimates the true degree to which
{\it TRACE} loops are over-dense \citep[cf.][]{wwm03}. Equation
(\ref{eq:n_eq}) is an idealized scaling law based on:  (1) an
approximate and somewhat outdated form for the radiative loss
function; (2) the assumption of no gravitational stratification; and
(3) the assumption of spatially uniform heating. The coefficient of
the scaling law should be treated with particular caution.
Furthermore, the ``actual" density determined from equation
(\ref{eq:density}) assumes a filling factor $f = 1$ and is therefore
a lower limit, but the equilibrium density determined from equation
(\ref{eq:n_eq}) does not depend on $f$. Despite these limitations,
$\Psi$ is a useful tool for evaluating the agreement between models
and observations.

\subsubsection{Intensity and Temperature}

We rejected the monolithic models on the basis of their 171
intensity properties, and we now examine whether the multi-strand
models fare any better.  We limit our discussion to the composite
models because we believe they are more realistic and, more
importantly, because they agree better with the observations.

The biggest failing of the monolithic models is their highly
structured intensity profile.  The problem is especially severe for
models with strong heating, which have localized bright emission
immediately flanking the cold condensation.  Multi-strand models
have a much more uniform appearance because they include many
unresolved condensations that are spread out along the loop.
Condensations tend to form in the upper two-thirds of the loop, at a
location that depends on the level of heating asymmetry and on
$\lambda$. The weaker the asymmetry, the closer to the apex they
form, with perfectly symmetric heating producing a condensation
right at the apex. Once formed, the condensations move downward
toward the footpoints. If all phases of the cycles are represented
in the strands, the entire loop will be filled in with bright
emission, including the lower legs, consistent with observations. It
is critical, however, that some of the strands have nearly symmetric
heating so that a dark gap is not present at the top.

\begin{figure}
\epsscale{.80} \plotone{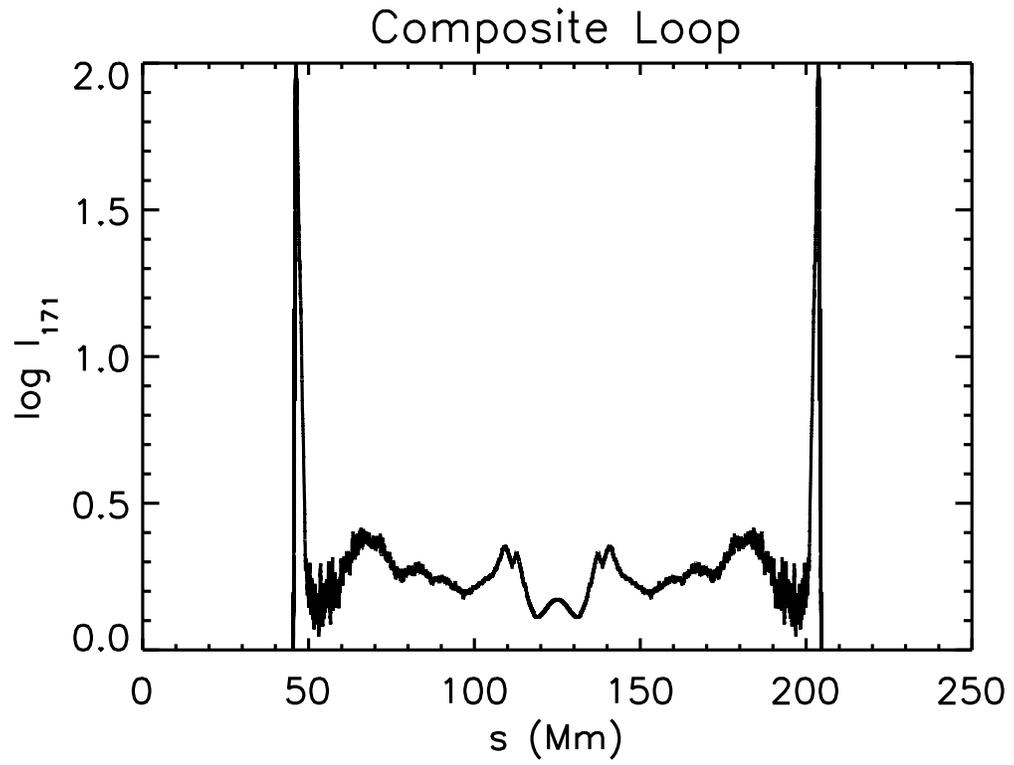} \caption{Logarithm of the {\it
TRACE} 171 intensity versus position for the composite loop with
strong heating. \label{fig:hot_comp_profile}}
\end{figure}

Figure \ref{fig:hot_comp_profile} shows the intensity profile for
the composite model with strong heating.  Except for the bright
spikes at the footpoints (note the logarithmic scale), the emission
is reasonably uniform.  Intensity variations along the loop are less
than a factor of 2 and would be smaller still if the bundle included
a larger variety of heating asymmetries.

\begin{figure}
\epsscale{.80} \plotone{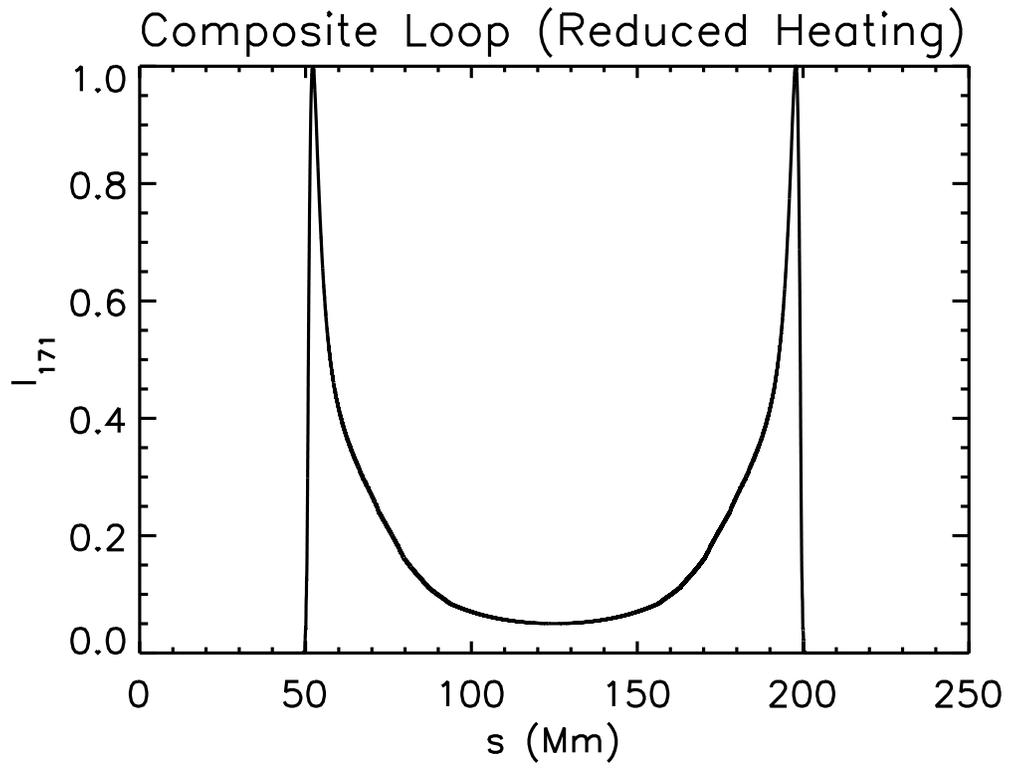} \caption{{\it TRACE} 171 intensity
versus position for the composite loop with weak heating.
\label{fig:warm_comp_profile}}
\end{figure}

Figure \ref{fig:warm_comp_profile} shows the 171 intensity profile
for the composite model with weak heating (linear scale). The
profile is very smooth, due largely to the fact that the individual
strands are reasonably uniform up to the time when the condensations
form. The profile is nonetheless inconsistent with observations
because the intensity decreases too rapidly with height.  The scale
height in the model corresponds to a hydrostatic loop at 1 MK,
whereas observed scale heights are super-hydrostatic by up to a
factor of 2.

\begin{figure}
\epsscale{.80} \plotone{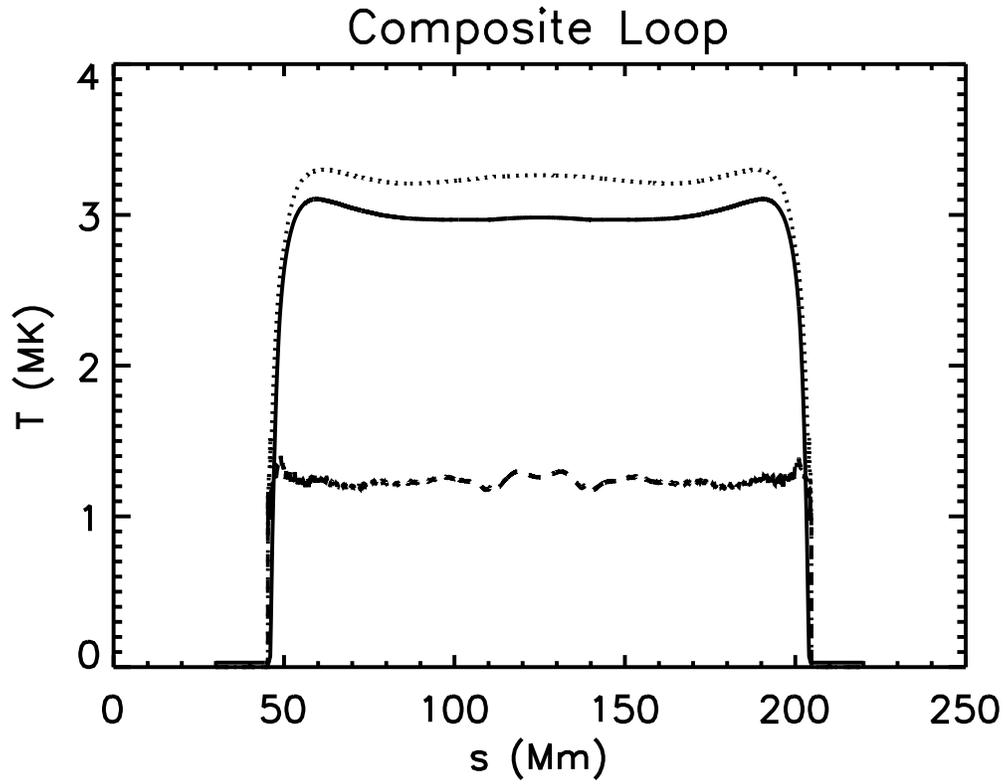} \caption{Temperature versus
position for the composite loop with strong heating: average of the
actual temperatures (solid), simulated {\it TRACE} temperature
(dashed), and simulated SXT temperature (dotted).
\label{fig:hot_comp_temp}}
\end{figure}

Figure \ref{fig:hot_comp_temp} shows three temperature profiles for
the strong heating composite model. The solid curve is the average
of the actual temperatures in the strands; the dashed curve is the
temperature that would be measured by {\it TRACE}; and the dotted
curve is the temperature that would be measured by SXT. The
temperature profiles are very flat, in excellent agreement with {\it
TRACE} observations and not inconsistent with SXT observations
\citep{kt96}. The composite model with weak heating also has a flat
{\it TRACE} temperature profile.  Its SXT profile is not relevant,
since the loop would be extremely faint in soft X-rays.

The multi-strand models presented here were obtained by temporally
averaging over two or more condensation cycles.  As such, they
represent very long-lived loops, inconsistent with observations.  We
could instead average over a portion of a cycle to obtain a shorter
lived loop, but then the intensity and temperature profiles would be
less uniform.  Averaging over a portion of the cycle corresponds to
condensations forming at roughly the same time in the different
strands.  If they form at the same time, they move together as a
group. The lower legs of such a loop would be dark in the early
stages of evolution, and the apex would be dark in the later stages,
neither of which agrees with observations. Whether it is possible to
build a loop that is both sufficiently short lived and sufficiently
uniform to match observed loops is a question that we examine in
more detail below.

\section{Discussion and Conclusions}

We have modeled monolithic and multi-strand loops undergoing thermal
nonequilibrium with the hope of reproducing the salient features of
observed loops, especially those seen in warm ($\sim 1$ MK)
emissions by instruments like {\it TRACE}.  A fundamental property
of these warm loops is their excess density compared to static
equilibrium. We find that many of our models can successfully
explain the observed densities.  Some can also explain the
unstructured intensities and flat temperature profiles that are
typically observed.  However, none of the models is able to
successfully reproduce all of the observed properties. The
monolithic models fail dramatically in that they have far too much
intensity structure. This is not a problem for the multi-strand
models, but these models, as presented, are far too long-lived. The
competing requirements of uniform intensity and short-to-modest
lifetime (1000--5000 s) are extremely difficult to satisfy.  It may
be possible to construct a model that satisfies both, but the
conditions are too contrived to be a plausible explanation for real
loops, as we now show.

It is instructive to briefly discuss the nanoflare storm concept
\citep{k09}, because it shares several common features with the
thermal nonequilibrium scenario we are now considering. In the
nanoflare case, each loop is envisioned to be a bundle of strands
that are heated impulsively at different times (but only once). At
any given moment, the many strands are in different stages of
cooling and therefore only some of them are detectable in the 171
channel. If the duration of the nanoflare storm (time delay between
the first and last nanoflare) is long compared to the lifetime of
each strand (duration of visibility), then the lifetime of the
entire loop bundle will be determined primarily by the duration of
the storm. If, on the other hand, the duration of the storm is short
compared to the lifetime of each strand, then the lifetime of the
bundle will be determined primarily by the lifetime of the strands.
It is straightforward to see that the loop lifetime is approximately
equal to the sum of the storm duration and the strand lifetime.

We can apply these same ideas to a bundle of strands undergoing
thermal nonequilibrium.  In place of the nanoflare storm duration,
we have the time delay $\Delta t$ between the formation of the first
and last condensations.  Just as there is only one nanoflare per
strand, there can be only one condensation (or condensation pair)
per strand, because the period of the cycles is considerably longer
than observed loop lifetimes. Let $\tau$ represent the time that
each strand is visible in the 171 channel. To reproduce the observed
loop lifetimes, $\Delta t$ must satisfy $\Delta t + \tau \approx$
1000--5000 s. Model strands with weak heating have $\tau > 5000$ and
can be immediately ruled out. Model strands with strong heating have
$\tau \approx$ 1000--2000 s (2000 s for the case with 90\% heating
asymmetry). Observed loop lifetimes can perhaps be reproduced if
$\Delta t \approx$ 0--4000 s.

The condition on $\Delta t$ is necessary but not sufficient. Loops
will have uniform intensity only if the strands are sufficiently out
of phase. There is a problem when $\Delta t$ is small because then
all of the strands are roughly in phase. The condensations form
together in the upper part of the loop and move together down the
leg.  The requirement of uniform brightness places a lower limit on
$\Delta t$ that is approximately the time it takes a single
condensation to traverse the entire half length of the loop. Only
then will one condensation appear near the apex at the same time
that another is about to disappear into the chromosphere. In the
simulation with 90\% heating asymmetry, the condensation takes
approximately 2000 s to traverse this distance. This is also how
long the strand is visible in 171. To have a uniform loop bundle
made from these strands implies a loop lifetime of at least $\Delta
t + \tau \approx 2000 + 2000 = 4000$ s. A majority of observed warm
loops are shorter lived than this. We conclude that they cannot be
explained by thermal nonequilibrium.

Even the longer lived loops are problematic.  To produce a
condensation, the heating in each strand must be steady or
quasi-steady for at least one cycle, which lasts approximately 2
hours. If the heating is steady for 2 hours, then it seems
reasonable to expect that it might remain steady for 4 hours, or
even longer. This would allow additional cycles to occur and the
loop to reappear multiple times. We can rule this out, however. As
shown in Table 1, strands with different heating parameters have
different cycle periods. The period also varies from one cycle to
the next for a given set of parameters (i.e., the cycles are only
quasi-periodic). Therefore, even if the phasing of the strands were
correct for the first appearance of the loop---itself a
challenge---it would not be correct for the second and subsequent
appearances.  To reproduce the observations, the heating must turn
on, remain steady for one full cycle, and then turn off before any
new condensations can form. This seems highly implausible.  We
conclude that thermal nonequilibrium is not a reasonable explanation
for any observed warm coronal loops, even those that are relatively
long-lived. Thermal nonequilibrium is also incapable of explaining
hot loops, since it cannot produce the under-dense conditions that
are characteristic of these loops.

An important implication of our results is that the dominant heating
mechanism in active regions cannot be both highly concentrated low
in the corona and steady or quasi-steady (slowly varying or
impulsive with a rapid cadence).  Active regions would look much
different if this were the case. Loops resembling our models---and
therefore unlike those observed---would be common. This claim must
be qualified with some caveats. It is acceptable for the heating to
decrease with height as long as the scale length is greater than
about 20\% of the loop half length. Only shorter scale lengths
produce thermal nonequilibrium. Even these short lengths might be
allowed if only one leg of the loop is heated, because then a steady
flow equilibrium can be established. It is unclear, however, whether
these steady equilibria can reproduce the excess densities,
intensity scale heights, and temperature profiles that are observed
\citep{pkm04,wetal02}.

Finally, we cannot exclude the possibility that thermal
nonequilibrium is occurring in the diffuse corona between loops. The
properties of this part of the corona are not well understand, and
evidence of thermal nonequilibrium might not be obvious if there is
a multitude of unresolved strands with random phasing of the cycles.
One consequence of many unresolved condensations would be the
absorption of the EUV radiation from below. Evidence of absorption
from unresolved cold material in the corona has been reported
\citep[e.g.,][]{so79,kg95}, but whether the quantities are
consistent with widespread thermal nonequilibrium has not yet been
investigated.

We close by emphasizing that thermal nonequilibrium is likely to
play an important role in the solar atmosphere under more limited
circumstances. It is almost certainly responsible for prominences
\citep{ak91,kak06}, and it may also explain ``catastrophic cooling
events," including coronal rain \citep{s01,detal04,obd07}.  During
these events, a cold ``blob" condenses out of the hot corona at the
top of a loop. It appears sequentially in the 195, 171, 1600, and
1216 channels of {\it TRACE}, which have maximum sensitivity at
temperatures of 1.5, 1.0, 0.1, and 0.02 MK, respectively. The blob
is visible only in the two coolest channels as it falls down the leg
at speeds of 20--100 km s$^{-1}$. Though fascinating, these events
are relatively uncommon.  The number of blobs observed at any one
time is much less than the number of warm loops (C. Schrijver, 2009,
private communication).

\citet{mph04} and \citet{as09} have suggested that these blobs are
formed by thermal nonequilibrium.  Our monolithic models with strong
heating have many similarities to the observations, including the
downward velocities, but key differences are not yet explained. The
observed time delay between the blob's appearance in the 171 and
1600 channels is more than twice what our models predict. More
significantly, our models predict 171 and 195 emission from the
transition regions that flank the blob as it falls, but such
emission is apparently not seen.  It is clear that more work is
needed before we fully understand the origin of catastrophic cooling
events.



\acknowledgments

This work was supported by the NASA Living With a Star program.  A
portion of it was performed while the authors were on the staff of
the Naval Research Laboratory.  We gratefully acknowledge useful
conversations with Roberto Lionello, Jon Linker, Yung Mok, Karel
Schrijver, and Daniele Spadaro.


\begin{table}[!ht]
\caption{Model Parameters}
\smallskip
\begin{center}
{\small
\begin{tabular}{ccccccccccc}
\tableline
\noalign{\smallskip}
$L$ & $Q_0$ & Asymmetry & Cycles & Period & $T_{TRACE}$ & $T_{SXT}$ & $\Psi$ & $\Psi_{TRACE}$ & $\Psi_{SXT}$\\

[Mm] & [erg cm$^{-3}$ s$^{-1}$] & [$\%$] & & [s] & [MK] & [MK] & & & \\
\noalign{\smallskip} \tableline \noalign{\smallskip}
75 & $8.0 \!\times\! 10^{-2} (strong)$ & 90 & 9 & 7330 & 1.30 & 3.35 & 5.07 & 9.69 & 3.00\\
75 & $8.0 \!\times\! 10^{-2} (strong)$ & 75 & 11 & 6370 & 1.31 & 3.16 & 4.09 & 6.79 & 3.15\\
75 & $8.0 \!\times\! 10^{-2} (strong)$ & 50 & 2 & 6800 & 1.36 & 3.15 & 3.56 & 4.74 & 3.02\\
75 & $8.0 \!\times\! 10^{-2} (strong)$ & Composite & 2-11 & & 1.24 & 3.24 &  4.26 & 8.88 & 3.01\\
75 & $8.0 \!\times\! 10^{-3} (weak)$ & 90 & 2 & 10,280 & 1.20 & 1.65 & 2.73 & 2.42 & 1.25\\
75 & $8.0 \!\times\! 10^{-3} (weak)$ & 75 & 2 & 10,280 & 1.19 & 1.47 & 4.69 & 2.53 & 1.52\\
75 & $8.0 \!\times\! 10^{-3} (weak)$ & 50 & 3 & 6370 & 1.06 & 1.11 & 5.18 & 3.66 & 2.99\\
75 & $8.0 \!\times\! 10^{-3} (weak)$ & Composite & 2-3 & & 1.14 & 1.51 & 4.05 & 2.85 & 1.45\\
150 & $2.0 \!\times\! 10^{-2} (strong)$ & 75 & 3 & 7700 & 1.22 & 3.37 & 6.62 & 11.34 & 3.10\\
37.5 & $3.2 \!\times\! 10^{-1} (strong)$ & 75 & 5 & 6110 & 1.29 & 3.28 & 2.90 & 4.98 & 2.35\\
\noalign{\smallskip} \tableline
\end{tabular}
}
\end{center}
\end{table}

\clearpage



\clearpage

\clearpage

\clearpage

\clearpage

\clearpage

\clearpage

\clearpage

\clearpage

\clearpage

\clearpage



\begin{thebibliography}{}

\bibitem[Antiochos \& Klimchuk(1991)]{ak91}
Antiochos, S. K., \& Klimchuk, J. A.  1991, ApJ, 378, 372

\bibitem[Antiochos et al.(1999)]{skaetal99}
Antiochos, S. K., MacNeice, P. J., Spicer, D. S.,\& Klimchuk, J. A.
1999, ApJ, 512, 985

\bibitem[Antolin \& Shibata(2009)]{as09}
Antolin, P., \& Shibata, K.  2009, \apj, submitted

\bibitem [Aschwanden, Nightingale, \& Alexander(2000)] {ana00}
Aschwanden, M.  J., Nightingale, R. W.,  \&  Alexander, D. 2000,
ApJ, 541, 1059

\bibitem [Aschwanden, Schrijver, \& Alexander(2001)] {asa01}
Aschwanden, M.  J., Schrijver, C. J.,  \&  Alexander, D. 2001, ApJ,
550, 1036

\bibitem[Aschwanden et al.(1999)]{aetal99}
Aschwanden, M.  J., Newmark, J. S., Delaboudiniere, J. P., Neupert,
W. M., Klimchuk, J. A., Gary, G. A., Portier-Fornazzi, F., \&
Zucker, A. 1999, ApJ, 515, 842

\bibitem[De Groof et al.(2004)]{detal04}
De Groof, A., Berghmans, D., van Driel-Gesztelyi, L., \& Poedts, S.
 2004, \aa, 415,1141

\bibitem[Kano \& Tsuneta(1996)]{kt96}
Kano, R. \& Tsuneta, S.  1996, \pasj, 48, 535

\bibitem[Karpen \& Antiochos(2008)]{ka08}
Karpen, J. T. \& Antiochos, S. K.  2008, \apj, 676, 658

\bibitem [Karpen et al.(2001)] {ketal01}
Karpen, J. T., Antiochos, S. K., Hohensee, M., Klimchuk, J. A., \&
MacNeice, P. J. 2001, ApJ(Lett), 553, L85

\bibitem [Karpen, Antiochos, \& Klimchuk(2006)] {kak06}
Karpen, J. T., Antiochos, S. K., \& Klimchuk, J. A. 2006, \apj, 637,
531

\bibitem [Karpen et al.(2003)] {ketal03}
Karpen, J. T., Antiochos, S. K., Klimchuk, J. A., \& MacNeice, P. J.
2003, \apj, 593, 1187

\bibitem [Karpen et al.(2005)] {ketal05}
Karpen, J. T., Antiochos, S. K., Tanner, S. E. M., \& DeVore, C. R.
2005, \apj, 635, 1319

\bibitem[Klimchuk(2000)]{k00}
Klimchuk, J. A.  2000, \solphys, 193, 53

\bibitem[Klimchuk(2006)]{k06}
Klimchuk, J. A.  2006, \solphys, 234, 41

\bibitem[Klimchuk(2009)]{k09} Klimchuk, J. A. 2009, in ASP Conf. Ser. XX,
Proceedings of the Second Hinode Science Meeting: Beyond
Discovery--Toward Understanding, eds. M. Cheung, B. Lites, T.
Magara, J. Mariska, \& K. Reeves (San Francisco: Astron. Soc.
Pacific)

\bibitem[Klimchuk \& Gary(1995)]{kg95}
Klimchuk, J. A., \& Gary, D. E.  1995, \apj, 448, 925

\bibitem[Klimchuk, Patsourakos, \& Cargill(2008)]{kpc08}
Klimchuk, J. A., Patsourakos, S., \& Cargill, P. J.  2008, ApJ, 682,
1351

\bibitem[Klimchuk \& Porter(1995)]{kp95}
Klimchuk, J. A., \& Porter, L. J.  1995, Nature, 377, 131

\bibitem[Lenz et al.(1999)]{letal99}
Lenz, D. D.,  DeLuca, E. E., Golub, L., Rosner, R., \& Bookbinder,
J. A. 1999, ApJ, 517, L15

\bibitem[L\'{o}pez Fuentes, D\'{e}moulin, \& Klimchuk(2008)]{ldk08}
L\'{o}pez Fuentes, M. C., D\'{e}moulin, P., \& Klimchuk, J. A. 2008,
\apj, 673, 586

\bibitem[L\'{o}pez Fuentes, Klimchuk, \& D\'{e}moulin(2006)]{lkd06}
L\'{o}pez Fuentes, M. C., Klimchuk, J. A., \& D\'{e}moulin, P. 2006,
\apj, 639, 459

\bibitem[L\'{o}pez Fuentes, Klimchuk, \& Mandrini(2007)]{lkm07}
L\'opez Fuentes, M. C., Klimchuk, J. A., \& Mandrini, C. H.  2007,
ApJ, 657, 1127

\bibitem[Mok et al.(2008)]{metal08}
Mok, Y., Miki\'{c}, Z., Lionello, R., \& Linker, J. A.  2008,
ApJ(Lett), 679, L161

\bibitem[M\"{u}ller, Hansteen, \& Peter(2003)]{mhp03}
M\"{u}ller, D. A. N., Hansteen, V. H., \& Peter, H. 2003, \aa, 411,
605

\bibitem[M\"{u}ller, Peter, \& Hansteen(2004)]{mph04}
M\"{u}ller, D. A. N., , Peter, H., \& Hansteen, V. H. 2004, \aa,
424, 289

\bibitem[O'Shea, Banerjee, \& Doyle(2007)]{obd07}
O'Shea, E., Banerjee, D., \& Doyle, J. G.  2007, \aa, 475, L25

\bibitem[Patsourakos, Klimchuk, \& MacNeice(2004)]{pkm04}
Patsourakos, S., Klimchuk, J. A., \& MacNeice, P. J.  2004, ApJ,
603, 322

\bibitem[Porter \& Klimchuk(1995)]{pk95}
Porter, L. J., \& Klimchuk, J. A.  1995, \apj, 454, 499

\bibitem[Reale \& Peres(2000)]{rp00}
Reale, F., \& Peres, G.  2000, ApJ(Lett), 528, L45

\bibitem[Rosner, Tucker, \& Vaiana(1978)]{rtv78}
Rosner, R., Tucker, W. H., \& Vaiana, G. S.  1978, ApJ, 220, 643

\bibitem[Schmahl \& Orrall(1979)]{so79}
Schmahl, E. J., \& Orrall, F. Q.  1979, ApJ(Lett), 231, L41

\bibitem[Schrijver(2001)]{s01}
Schriver, C. J.  2001, \solphys, 198, 325

\bibitem[Susino et al.(2010)]{setal10}
Susino, R., Lanzafama, A. C., Lanza, A. F., \& Spadaro, D.  2010,
\apj, 709, 499

\bibitem[Testa, Peres, \& Reale(2005)]{tpr05}
Testa, P., Peres, G., \& Reale, F. 2005, \apj, 622, 695

\bibitem [Ugarte-Urra, Warren, \& Brooks(2009)]{uwb09}
Ugarte-Urra, I., Warren, H. P., \& Brooks, D. H.  2009, ApJ, 695,
642

\bibitem [Ugarte-Urra, Winebarger, \& Warren(2006)]{uww06}
Ugarte-Urra, I., Winebarger, A. R., \& Warren, H. P. 2006, ApJ, 643,
1245

\bibitem[Vesecky, Antiochos, \& Underwood(1979)]{vau79}
Vesecky, J. F., Antiochos, S. K., \& Underwood, J. H. 1979, \apj,
233, 987

\bibitem[Winebarger \& Warren(2005)]{ww05}
Winebarger, A. R., \& Warren, H. P.  2005, ApJ, 626, 543

\bibitem [Winebarger, Warren, \& Mariska(2003)]{wwm03}
Winebarger, A. R., Warren, H. P., \& Mariska, J. T. 2003, ApJ, 587,
439

\bibitem [Winebarger, Warren, \& Seaton(2003)]{wws03}
Winebarger, A. R., Warren, H. P., \& Seaton, D. B. 2003, ApJ, 593,
1164

\bibitem[Winebarger et al. (2002)]{wetal02}
Winebarger, A. R., \& Warren, H., van Ballegooijen, A., DeLuca, E.
E., \& Golub, L.  2002, ApJ, 567, L89

\end{thebibliography}
\end{document}